# Mobility Enhancement of Patients Body Monitoring based on WBAN with Multipath Routing

Yasna Ghanbari Birgani
Industrial Engineering Department
Tarbiat Modares University
Tehran, Iran
yasna.ghanbari@modares.ac.ir

Nastooh Taheri Javan
Computer Engineering Department
Amirkabir University of Technology
Tehran, Iran
nastooh@aut.ac.ir

Mohsen Tourani
Science Department
K. N. Toosi University of Technology
Tehran, Iran
mh_tourani@yahoo.com

***Abstract*—One of the promising applications of wireless sensor networks (WSNs) is monitoring of the human body for health concerns. For this purpose, a large number of small sensors are implanted in the human body. These sensors altogether provide a network of wireless sensors (WBANs) and monitor the vital signs and signals of the human body; these sensors will then send this information to the doctor. The most important application of the WBAN is the implementation of the monitoring network for patient safety in the hospital environment. In this case, supporting patients' mobility is one of the basic needs, which has been underestimated in recent studies. The problem that involves providing the required energy for the units used in this type of network is challenging; for this reason, sent/ received units with very low power consumption and with a very small radius are used in order to save energy. The resulting small sending range, leads to the lack of support for patients' mobility. In this paper, the AD HOC mode is suggested for use to establish a network and a multi-path routing algorithm, for the purpose of importing patients' mobility in hospital setting. The results of the simulation show that in addition to supporting patients' mobility, the use of the proposed idea instead of previously presented protocols, reduces delays in data transmission and energy consumption; and it also increases the delivery rate depending on the destination and the lifetime of the network, while on the other hand, it increases routing overhead.**

***Keywords— WBAN; Health Monitoring; Patient Mobolity; Routing Algorithms***

## I. INTRODUCTION

Recent advancements in wireless sensor networks (WSNs) have created new applications in wireless medical environments, whether the environment is a hospital room, a bedroom, or the inside of an ambulance. In this environment, there is a WSN that measures the physiological parameters of the body. It then transfers this data to an access point (AP) near the body. After this, the data from these APs via another network, such as the wireless Local Area Network (WLAN), is sent to the databases available to the medical staff [1]. The main advantages of measuring the parameters through wireless equipment are simplicity and facility.

WBANs are networks placed on the human body that are composed of multiple wireless sensor nodes used for monitoring one or more physiological parameters to make the remote monitoring of the environment possible [2]. Each node senses one or more physiological signal. The obtained data collected by the sensors is transmitted to the central unit. This device can send the data in a raw form or after the first processing to the physician.

One of the main problems encountered in order to implement the practical WBANs in real environments (such as a large hospital), is the supporting of patients' mobility [4]. This subject is important, because in order to reduce the energy consumption, the sent/ received units with very low energy consumption, are used in these networks. For this reason, the sending and receiving range for each device is very small, so that the environment coverage in this network is equal to the patient's mobility. In many of the available implementations (e. g. [5]), patient mobility is almost ignored; thus, the available mobility area for the patient will be considered his/her bed. This may be the case for certain patients, who cannot move in any way, and it can be acceptable to some extent; but, in general this limited coverage cannot answer the needs of other patients in other sectors.

To support patient mobility in general, three strategies have been conceived; the first solution is as follows: use the existing communication infrastructures, such as GSM, and apply equipment such as mobile phones; the second solution suggested the following: use a proprietary platform for WBAN and cover the general environment with multiple APs; and, the third approach is as follows: use the AD HOC mode.

The first solution can be applied on data communication power infrastructures for medical use. One common approach in this field is to use GSM and cell phones. In this strategy, with the help of GPRS protocol, the data collected from the patient and through the mobile phone network is sent to the physician. The biggest disadvantage of this method is the dependence of the health monitoring system and its ability to efficiently monitor the human body on the performance of cell phones. Actually, in this case, the data transmission delay will be high which can cause difficulties in critical medical applications. On the other hand, the system unavailability due to network congestion or common problems of this type can cause serious problems.

Another general idea is to use a specific substrate for this application. This means that a particular wireless network for the purpose of sending medical data should be created; thus, in this case, by considering the small sending range of these machines (which is mainly due to a restricted energy source and the high energy consumption of the machines) the mobility issue (or in other words the environmental coverage) should be solved in some ways. Two approaches are used for this purpose; in the first case, the whole environment could be covered with multiple APs. For this purpose, a hospital may need tens or even hundreds of APs to cover the entire perimeter which raises the cost of implementing this idea. Despite this cost, some benefits such as single hop data sending and low end to end delay will be gained [6, 7].

In contrast to this solution, there is the AD HOC mode. In this method, in addition to the task of collecting vital patient signs, machines also act as routers, which shift the data to other stations. Thus, patients together form a mobile ad hoc network. The main advantage of this method is the fast and low cost of the network setup; on the other hand, because of multi-hop transmissions, the level of lag (delayed transmission) post in this way will be somewhat higher. On the other hand, there is more or less the possibility of breaking the network, which increases the rate of packet loss.

In this paper, a multi-path routing protocol has been presented for WBAN networks in a multi-Hop position. In this position, it is assumed that other nodes in AD HOC mode start routing and shifting the data of others. In this way, stations in the process of route discovery, find several different routes to the destination node and try to send their data through several routes simultaneously, while data is being transferred. Thus, reliability is increased and the dynamics of topology and the detachment of links have much less of an impact on packet loss rate.

The rest of this paper is composed of the following sections; in the second section, previously published related articles and proposed methods are reviewed and in the third section, the details of the proposed idea are examined. The fourth section is dedicated to providing simulation results and the fifth section concludes the paper.

## II. RELATED WORKS

In WBANs, in order to support patients' mobility by using the AD HOC mode, a great deal of investigations have been undertaken.

In [8], with an approach to reduction of energy consumption a protocol has been presented to support multi-hop transmission in the WBANs. In this approach, in order to confront the changes in the network topology (due to patients' mobility), according to the needs of stations, changes in topology can be used immediately by the sending stations. In this protocol, a method based on a TDMA used to control access to the channel and a method based on gossiping for routing have been used. The authors in this article have provided a transmission ability method which is based on the measurement of the quality of the established links, so they can always reach the minimum energy consumption.

The idea of multi-step in WBAN networks is discussed in [9 and 10]. In this research, the changes necessary in the channel access layer of the networks are discussed. The obvious thing this research lacks is a consideration for the movement of the stations.

In [11], the authors have listed the mobility requirements of WBANs. They believe that WBANs are relatively small networks that the quality of the communication channel is severely altered in them. On the other hand, through the mobility of stations (patients), the network topology becomes highly dynamic; yet, the network must always remain attached. The solutions should be proposed in a way that satisfies the requirements expressed in this research.

In recent years, several routing protocols have been proposed for use in WBANs. In [12], the authors presented a distributed routing approach based on projections to enhance the security and reliability. In this method, each node estimates the condition of links in the future based on the quality of links in the past; and based on this estimation, a respective node chooses amongst its neighbors for sending. In this case, when a number of links (neighbors) exist to send data, this algorithm selects a higher quality link which is planned.

In [13], the authors have presented a multi-step routing algorithm for WBANs based on patients' mobility. In this way, they actually preserve the history of the mobility of the nodes so that they are able to estimate possible topological changes according to it. Their idea has led to some improvements in the reduction of latency in comparison to some existing methods.

Through the implementation of WBAN in hospital environments (sometimes referred to as Hospital BAN) some research has been done. In [14] the authors implemented the idea of the relationship between WBANs in the hospital setting; but, the reliability and quality of service parameters are not considered. In [15], with the aim of increasing the reliability in communication, a routing idea aware of the quality of service has been provided. In this case, the authors divided the apparatuses used in these networks in terms of energy into three categories. The first group is the equipment of nursing stations that are always plugged in and do not have a shortage of energy. The second set is the medical monitoring coordinators that have interchangeable batteries; and, the third category is the BAN coordinators that have very little energy.

On the other hand, multi-path routing, sometimes called traffic dispersion, has been one of the most important current directions in the area of routing in wireless and mobile networks [16, 17, 18]. Multi-path routing protocols will find multiple routes which have the advantages of easy recovery from a route failure and being more reliable and robust. Moreover, the source can select the best one among multiple available routes [19, 20].

As it was mentioned in the second section, in order to support patients' mobility in the hospital, a little research has been done using the AD HOC mode [9, 10]. In general, there are 2 inherent problems encountered when using the AD HOC mode for data transmission in these networks. The first issue raises the delay in data collection caused by the multi-step transmission. The second issue is the increase in the packet loss rate which is due to the deterioration and loss of links.

In the proposed RMPR (Reliable Multi-Path Routing) algorithm, there has been an attempt to correct routing and the route discovery procedure, and to discover and communicate multiple appropriate paths between each node to the destination (typically sink node). Each node will then try to send its maximum data through three paths to the destination. The ultimate goal is to improve the parameters of delay and the packet loss rate, so that in this rout, some costs like the increased routing overhead will be paid for. In summary, we can say that in order to reduce the delay, the multi-path idea has been used, and to reduce the loss rate of packets, the reliability of routes are measured and considered.

The routing algorithm RMPR can be reviewed in three main phases:

- Phase I: Route Discovery
- Phase II: Route Maintenance
- Phase III: Sending data

### A. Route Discovery

The proposed algorithm RMPR has been based on on-demand routing algorithms. In on-demand algorithms, when the source has data to send to a specific destination, but does not have the path to the destination, the route discovery process is executed. In this case, the source sends a route request packet (RREQ) to all its neighbors by flooding the way. Each packet of the route request stores the list of nodes it has traversed in its header. Since the source spreads packets universally, it is likely that the destination will receive several packets of the route request from different directions. In this case, the destination can send some route reply packets (RREPs) through multiple paths to the source. In the proposed algorithm, all nodes should store a table of the seen RREQs (in order to avoid the creation of loops), so that in each table, the specifications for received RREQs are recorded by each node. In this case, when an intermediate node receives a RREQ, it searches its specifications in its own seen RREQ table; and, if the packet is new, first, the specifications will be listed in the seen RREQ table, then it will send the packet to its neighbors. With this approach, only the Node-Disjoint paths at the source will be discovered. The destination node returns the route reply packets (RREPs) after receiving the RREQs through the reverse path of each RREQ to the source.

AS shown in Figure 1, node S which is the source node, sends the RREQ through a flooding method to its neighbors that are A, B and C nodes. Each of these three nodes first check their seen RREQ table to verify if the application packet will not be repeated. They then start to calculate the Strength factor. For example, the ratio of the received power of node A from node S is 0.8. Node A updates the Strength factor of the field. The source node (S) initializes this field with number 1. Node A calculates the Strength factor of the upstream packet which is 0. 8, as soon as it receives the request packet and multiplies the Strength Factor which is present in the packet. So far along this path, the Strength Factor is 0.8. This process continues until the request packet reaches the destination which is node D. At last, the Strength factor of S-A-D paths will be 0.56. For the other two discovered paths S-B-D and S-C-D, the Strength Factor is calculated in the same way.

The destination which is node D, creates the corresponding response (or Reply) packets with discovered paths that are three paths; S-A-D, S-B-D and S-C-D; and, it copies the Strength Factor field to define it, and sends the packets to the source. The source will identify the path with the highest Strength Factor most suitable for sending.

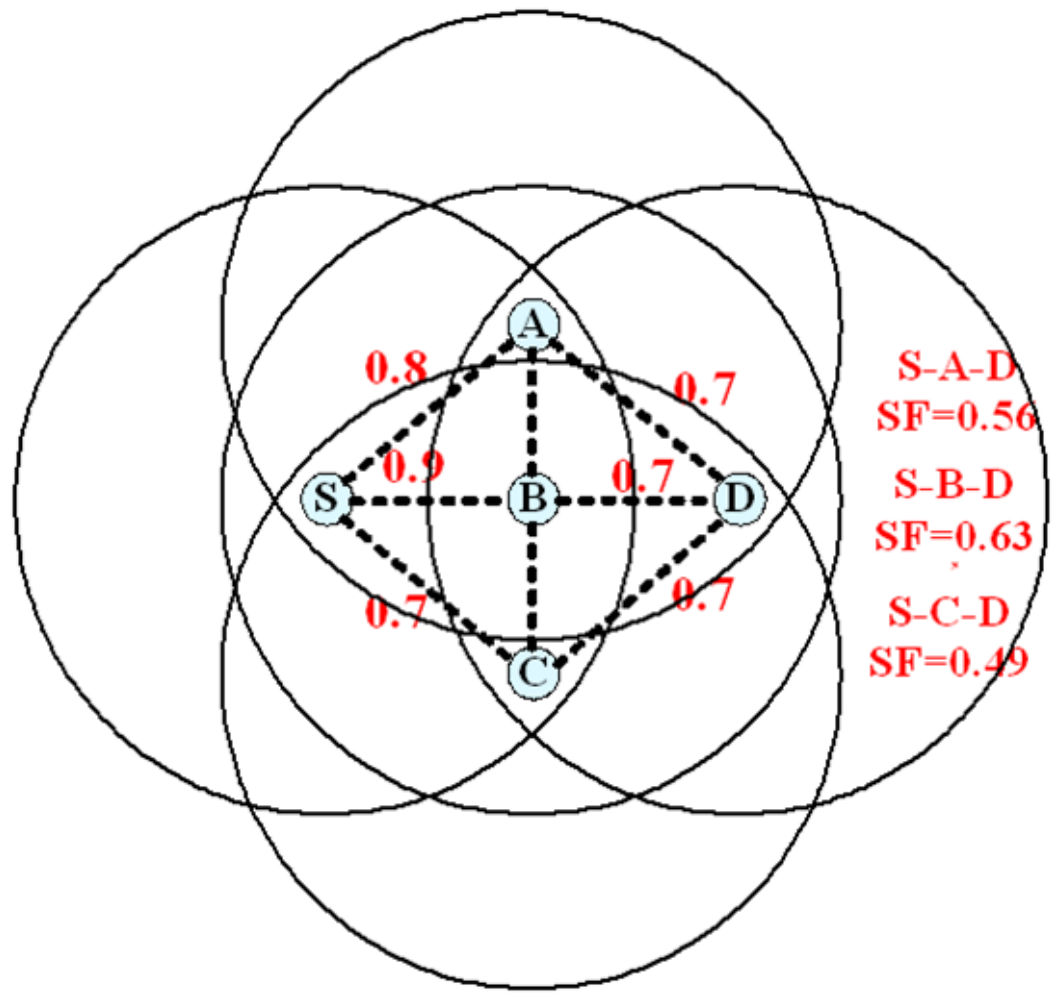

Figure 1. Route Discovery phase

### B. Route Maintenance

After route discovery in the first phase, it is likely that over time and with patients' mobility, the quality of routing paths will drop or that routes will become detached. On the other hand, the source faces the following fundamental concern: the issue of determining among the discovered routes in the first phase, the route which has the highest priority to send data.

For this purpose, the destination through a knowledge of the paths, sends a packet described as a "packet of measurement” to the source. During the movement of the packet of measurement, each node updates the present parameter "Strength Factor" in the packet of measurement. For this purpose, the destination, by sending a packet of measurement, initializes the parameter of Strength Factor with number 1 in that packet. Each node based on the received power signal from the superior node, now updates the amount of Strength Factor available in the packet of measurement. For this purpose, each node calculates the Strength Factor parameter (SF) as follows:

$$SF = \frac{S_{XY} - S_{min}}{S_{XY}} \quad (1)$$

In (1) $S_{XY}$ is equal to the Signal power that node X received from node Y; and, in fact, $S_{min}$ is acceptable as the lower threshold for a received signal strength. So, SF is indeed a normalized display for the signal strength.

As it was mentioned, the destination sends the packet of measurement with the initial measurement equal to one for the SF field to the source. Each node by receiving the packet of measurement from its upstream node, calculates the related Strength Factor and multiplies it to the amount of the available Strength Factor in the packet of measurement. In fact, when the packet of measurement reaches the source, the amount of the present Strength Factor is equal to the amount of the total Strength Factor (the multiplied Strength Factor of the links). The source can always identify and use the most powerful routes with the help of this parameter: each route that has the higher Strength Factor is more suitable for transmission.

Whenever the number of powerful routes for a source is reduced from a threshold, the source can replace the newer routes by re-running the route discovery phase.

As it is seen in Figure (2), the destination node (node D), in considering the 3 reply packets sent to the source, provides the following 3 route measurement packets: S-A-D, S-B-D and S-C-D; and, it sends them to the source. The field of the present Strength Factor in the measurement packets is updated. For example, after initializing an SF field by the destination (node D), when the packet reaches node A, this node measures the received signal power from its upstream node; the node in question is node D which has a signal power of 0.7 and it multiplies this to the previous amount. This process continues until the measurement packet reaches the source. The Strength Factor field amount in the packet of measurement related to the S-A-D route will be 0.49. Thus, the source will be able to choose the path with the higher reliability at any moment.

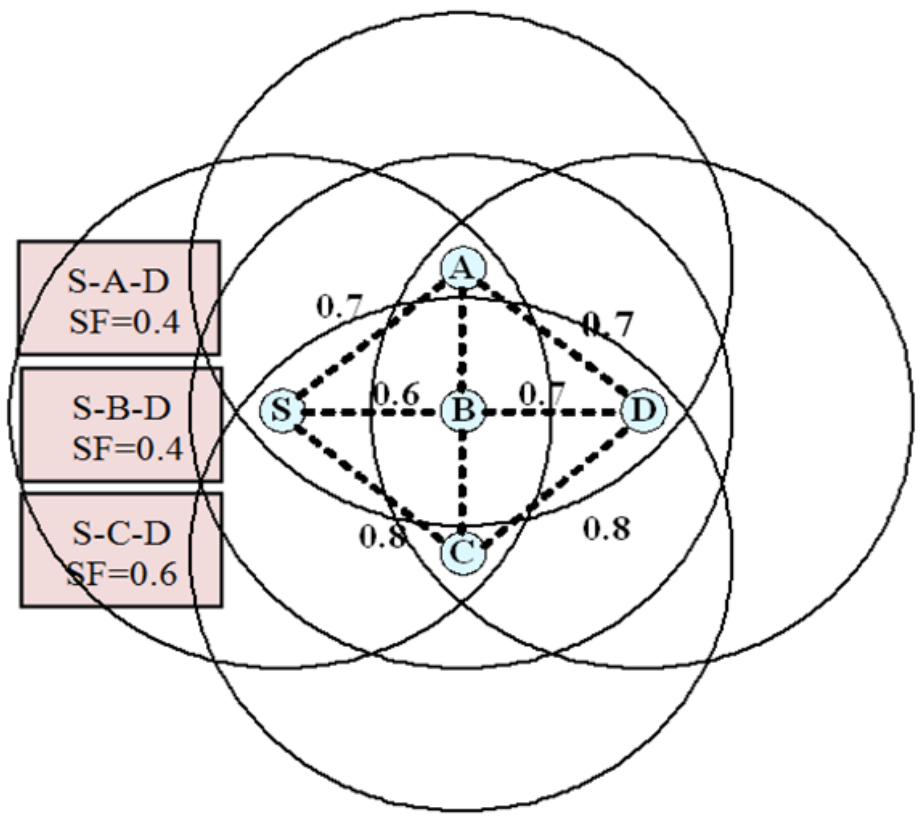

Figure 2. Route maintenance phase

### C. Data Sending

The source node always tries to send its data to the destination via two routes simultaneously. For this purpose, it distributes the sending packets in an alternating way between two routes. The key point in the proposed idea is to use the routes with a higher reliability in order to send data. For this purpose, the source tries to discover the most powerful routes in terms of time efficiency with the help of retention (path maintenance) mechanism.

## III. EVALUATION AND RESULTS

In this section, the results of the simulation are presented. We have used the OMNet++ software and have added the Castalia to it in order to conduct the simulation. For this purpose, the results of the following 3 ideas have been compared:

- The proposed algorithm RMPR.
- The referenced idea [14] as the first algorithm.
- The referenced idea [15] as the second algorithm.

All three cases are within the multi-hop sending category and the AD HOC mode; and, they cannot be compared to non-AD HOC methods which are due to the inherent differences between these methods.

In [14], the method in the form of the first algorithm is referenced; and, the available devices in the network are divided into three categories that consider energy. The first category is the nursing equipment that is always plugged in and does not have a shortage of energy. The second category is the display nodes that have rechargeable batteries; and, the third category covers nodes that are in the patients' bodies, that collect the vital signs, and which are not rechargeable. In this method, three types of nodes are predicted and the network has a lot of limitations and complexities when it comes to implementation. In this way, the reliability of routes would be measured.

Method [15], the second algorithm, is composed of clusters that include a number of patients that each cluster has a cluster head. This method requires a special node for cluster heads with adjustable sending ranges used to reach the APs.

In this simulation, 200 nodes (patients) were used in an environment with the following dimensions: 200 m × 200 m. In this case, the transmission radius that has been considered for each node is 15 m; and, in order to simplify the implementation, the Random Way Point is used as the mobility mode for patients. In this model, each node selects a point as a destination accidentally, then with a speed between a minimum and maximum speed, the respective node will moves towards the destination. Once the destination is reached, it stays in the same spot for a period of time which has been marked as the pause time, and then it repeats the same procedure. In all the simulations, we have considered the pause time as equal to three seconds. On the other hand, in some of the scenarios, in order to naturalize patients' mobility, with the help of Castalia, some of the mobility restrictions can be applied to the nodes. We also have randomly created a CBR traffic for nodes. In this simulation, the nodes for sending and receiving data, use the standard radio model RADIO-ACCNOISE; and, in order to access the channel, the standard IEEE 802.15 is used by default. Each simulation time is considered as 600 seconds and each of the recorded results is the average of 15 times the scenario running.

The overall goal of the simulation scenario is to check the behavior of the algorithms against the increase of patients' mobility. For this purpose, the maximum speed of each patient in the Random Way Point mobility model, increases up to 50 meters per minute (3 km/h). In this way, it will try to estimate the impact of patients' mobility on the parameters of assessment. The results of the implementation of simulations are presented in the following text.

### A. End-to-end delay

In figure (3), the three considered frameworks are compared to each other in terms of the average end-to-end delay for data packets received by the sink node. In Figure (3), patients' mobility has increased gradually. As shown in this figure, in the proposed PMPR algorithm, the average of the end-to-end delay is less than that of the other two algorithms; this is the result of PMPR using two routes to send data simultaneously. On the other hand, in all three scenarios where there is an increase in patients' mobility, and increase in the rate of lost connections, and the need to resend packets or reinitialize route discovery, the end-to-end delay will be increased greatly.

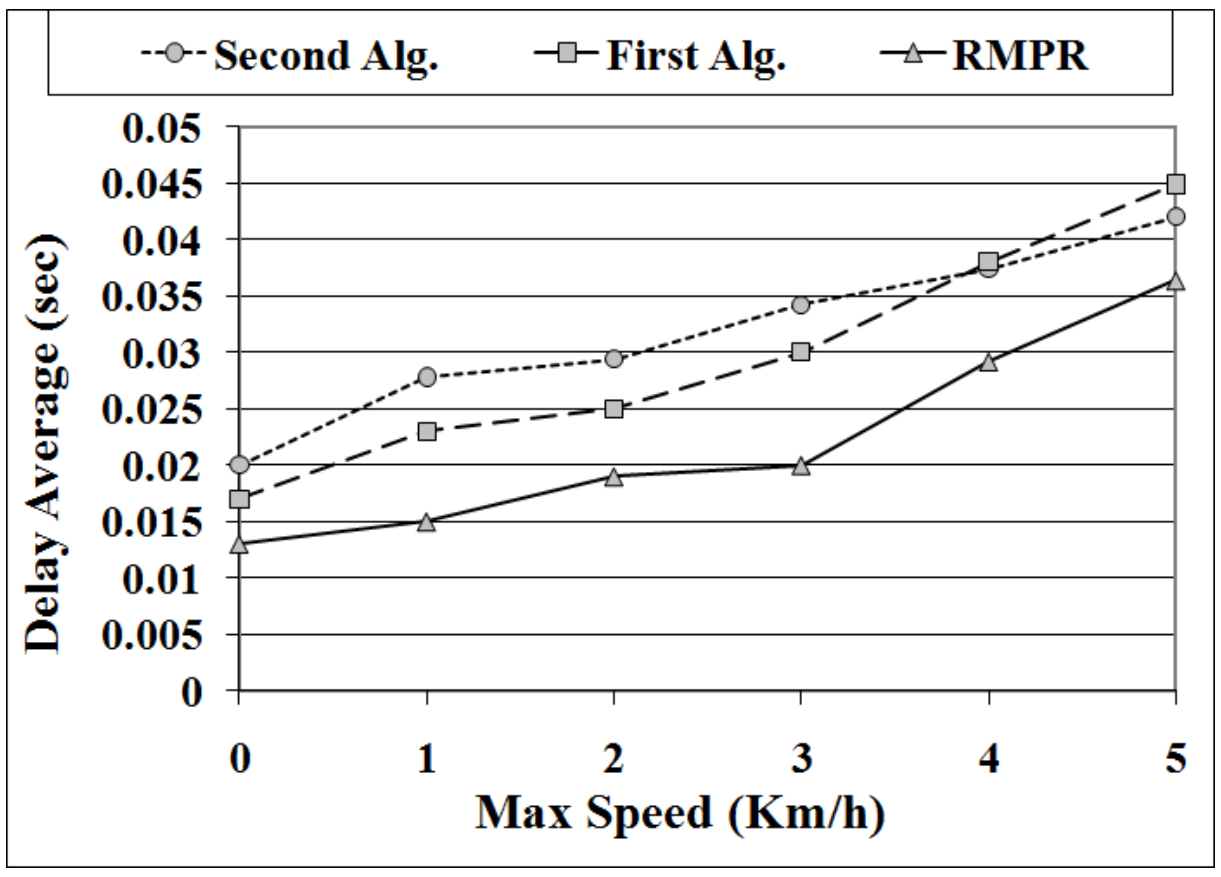

Figure 3. End-to-End Delay

## *B. Packet Delivery Ratio*

In Figure (4), the aforementioned algorithms are compared to each other by considering the rate of loss of packets. In these scenarios, with the increase in patients' mobility and the loss of links, the inevitable result is the loss of some data packets. In Figure (4), the percent of packets delivered to the destination is shown. Considering this figure, it demonstrates that in the proposed algorithm the situation is much more favorable due to the discovery and use of routes with high reliability.

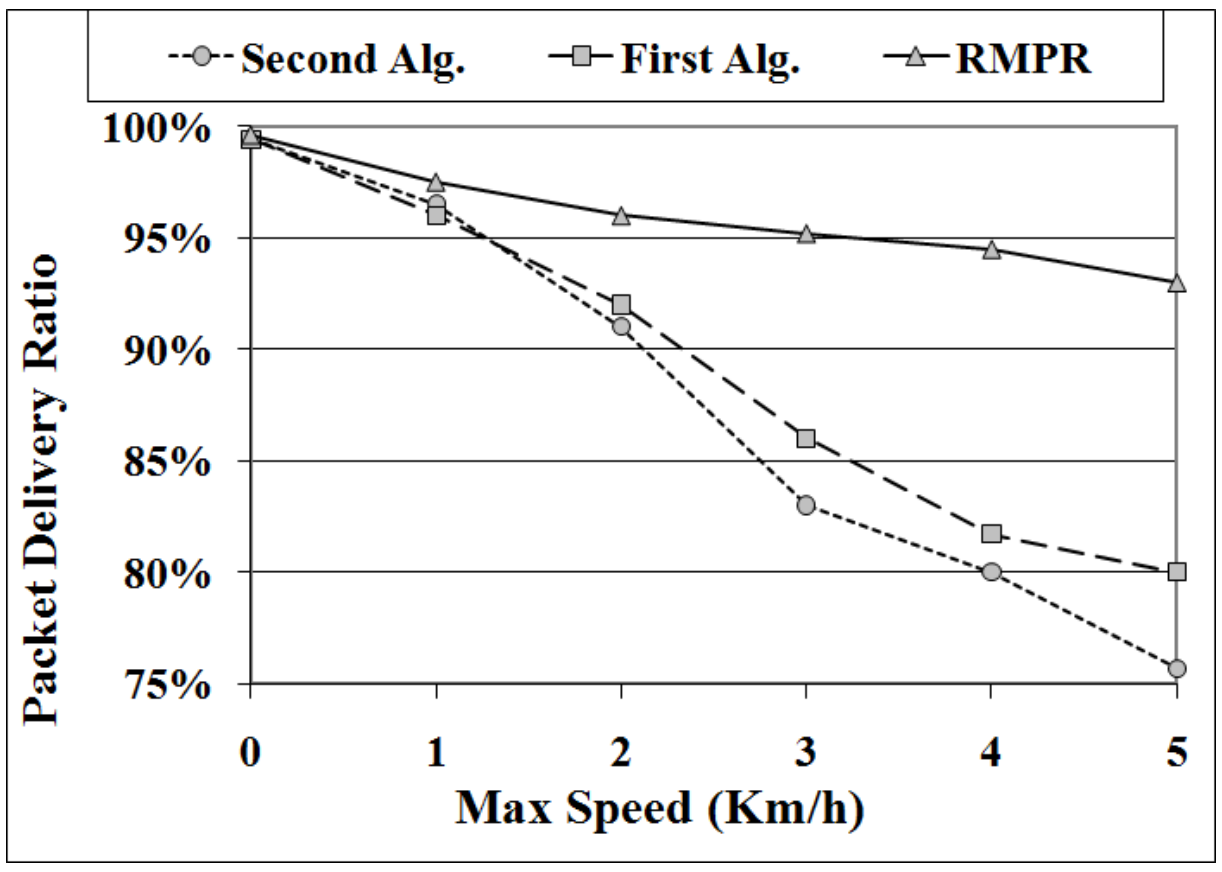

Figure 4. Packet delivery Ratio

## *C. Energy Consumption*

In Figure (5), the energy consumption is surveyed by the change in patients' mobility. As it is observed in Figure (5), with the increase in patients' mobility, the consumed energy increases. The increase in patients' mobility causes the loss of routes and the need to re-run the route discovery process that in turn results in an increase in energy consumption. In the proposed method, the energy consumption is lower than in the other two methods because of the reduction in the rate of packet loss; and, as a result, there is a reduction in the resending and distribution of energy consumption between two paths.

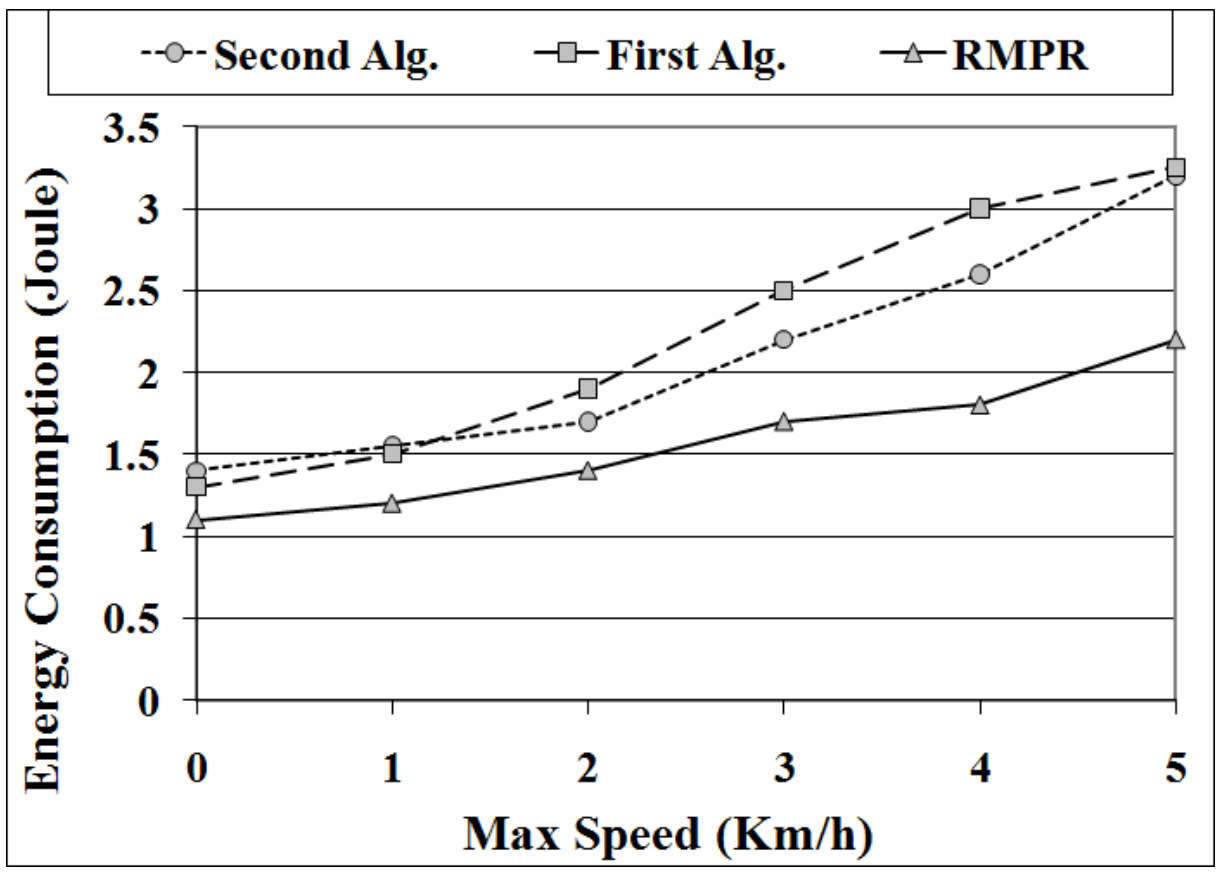

Figure 5. Energy Consumption vs. Patients' Mobility

## *D. Network Lifetime*

In Figure (6), the impact of patients' mobility on the lifetime of the network is shown. As it is obvious in this figure, the first blackout occurs more quickly because of the higher consumption of energy that occurs with the increase in the mobility of nodes. PMPR Algorithm has a longer lifetime in comparison with the two multi-hop methods due to its method of sending packets through paths with higher reliability.

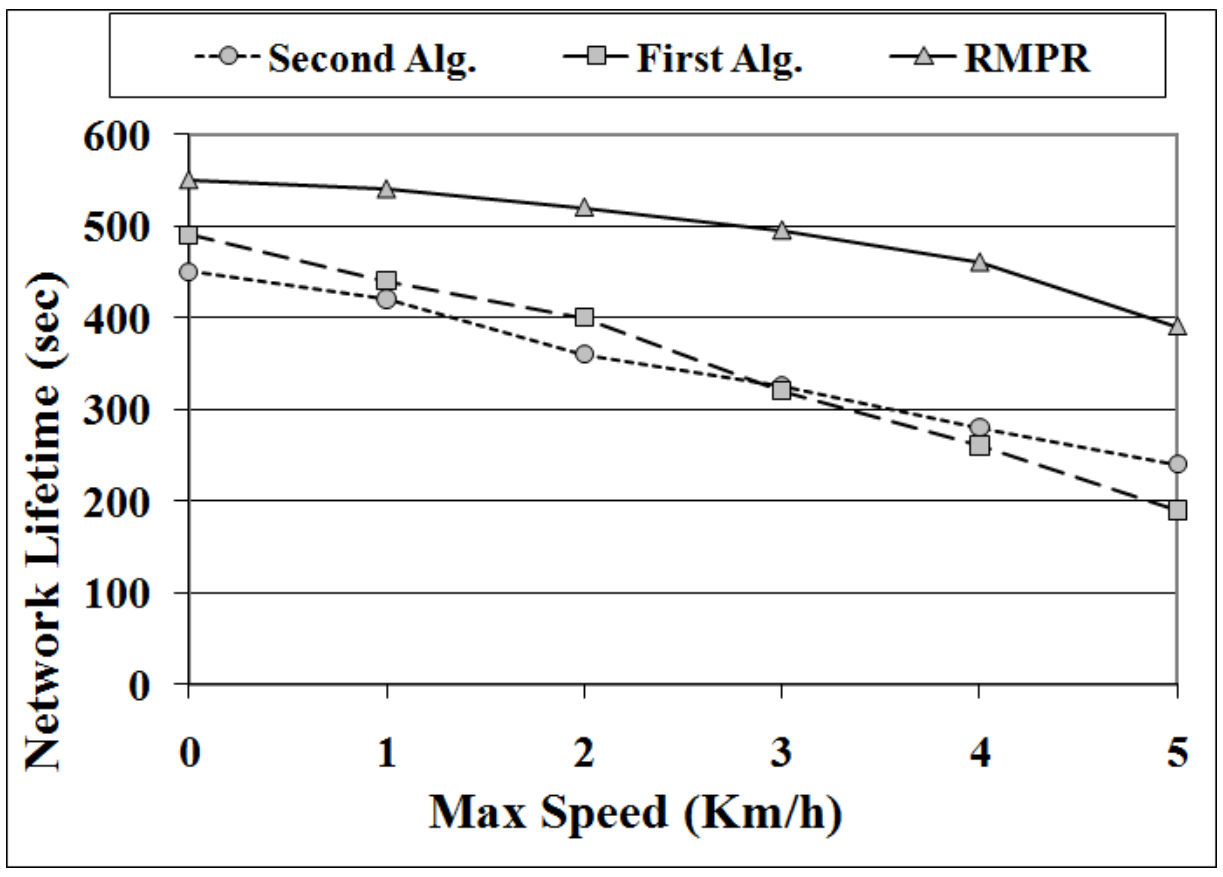

Figure 6. Network lifetime vs. patients' mobility

## *E. Routing Overhead*

In Figure (7), the amount of routing overhead is determined for the three algorithms. Here the routing overhead means the number of routing packets sent per packet. As it is shown in Figure (7), the proposed algorithm PMPR has a higher overhead than the other two methods that have pretty acceptable level of overhead; this discrepancy is due to the use of multiple packets in the route discovery and the choosing process.

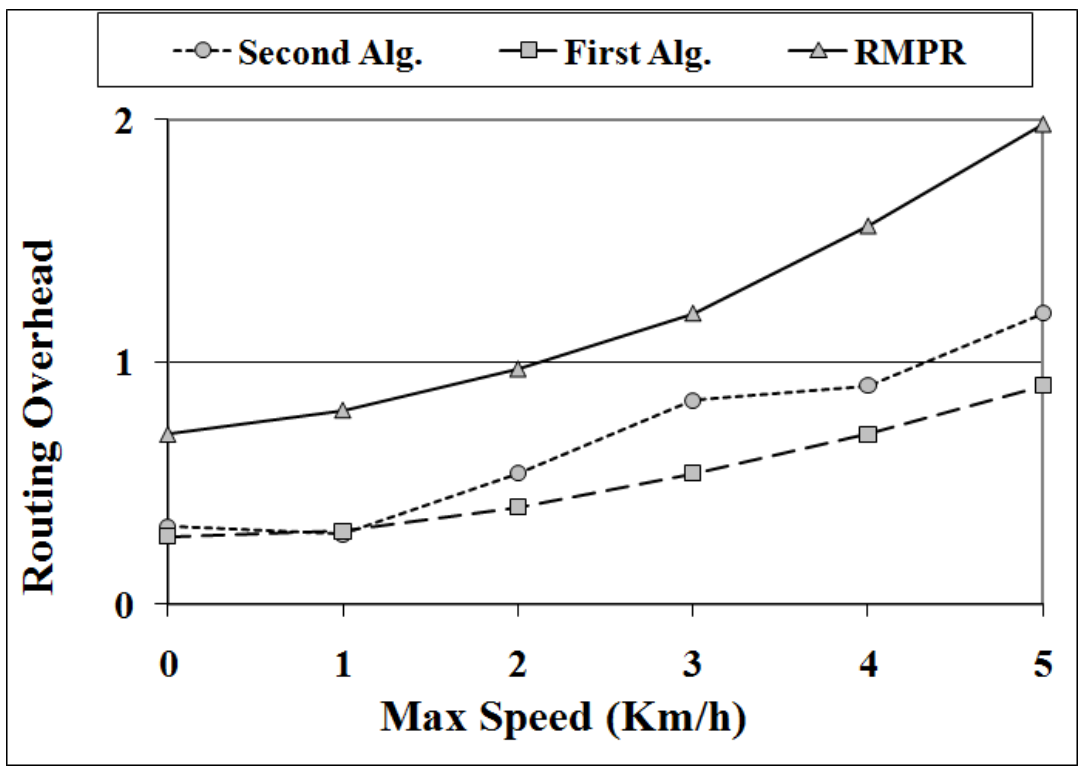

Figure 7. Routing overhead

## IV. CONCLUSION

The use of wireless sensor networks (WSNs) to monitor patients' health will be inevitable in tele-medicine in the upcoming years. Today, this newfangled technology faces numerous challenges. One of the challenges ahead is the support of patients' mobility in their respective environments (e.g., the environment of hospitals). Many of the present implementations, and much of the researches have ignored the issue of patients' mobility. On the other hand, some studies have incorporated the AD-HOC mode in these networks. But in this study, because of multi-Hop transmission, the sending delay increases and the probability of packet loss also increases. In this paper, a reliable multi-path routing algorithm (RMPR) is proposed to support the mobility of patients. The proposed algorithm after discovering some routes, tries to send data, if possible, through a maximum number of three paths simultaneously. The results of the simulation show some improvements in the reduction of end-to-end delay and the reduction of the packet loss rate over after present techniques. On the other hand, we witness the increase in overhead of the routing data in the proposed algorithm.